# Domain – wall – induced magnetoresistance in pseudo spin – valve/superconductor hybrid structures

A. K. Suszka[1*], F. S. Bergeret[2,3] and A. Berger[1]

[1] CIC nanoGUNE Consolider, E-20018, Tolosa Hiribidea 76, Donostia-San Sebastian, Spain

[2] Centro de Física de Materiales (CFM-MPC) Centro Mixto, CSIC-UPV/EHU, E-20018 Donostia-San Sebastian, Spain

[3] Donostia International Physics Center (DIPC), Manuel de Lardizabal 4, E-20018 Donostia-San Sebastian, Spain

We have studied the interaction between magnetism and superconductivity in a pseudo-spin-valve structure consisting of a Co/Cu/Py/Nb layer sequence. We are able to control the magnetization reversal process and monitor it by means of the giant magnetoresistance effect during transport measurements. By placing the superconducting Nb-film on the top of the permalloy (Py) electrode instead of putting it in between the two ferromagnets, we minimize the influence of spin scattering or spin accumulation onto the transport properties of Nb. Magnetotransport data reveal clear evidence that the stray fields of domain walls (DWs) in the pseudo-spin-valve influence the emerging superconductivity close to the transition temperature by the occurrence of peak-like features in the magneto-resistance characteristic. Direct comparison with magnetometry data shows that the resistance peaks occur exactly at the magnetization reversal fields of the Co and Py layers, where DWs are generated. For temperatures near the superconducting transition the amplitude of the DW-induced magnetoresistance increases with decreasing temperature, reaching values far beyond the size of the giant magnetoresistive response of our structure in the normal state.

# I. INTRODUCTION

Conventional superconductivity and ferromagnetism are in principle incompatible phenomena in bulk materials. While the ground state of a superconductor relies on Cooper pairs in the singlet spin state, ferromagnetic exchange tends to align the spin of electrons. These antagonistic phenomena may however coexist in hybrid superconductor-ferromagnet (SC-FM) thin film structures and the interplay between them leads to many interesting effects that have been studied during the last years [1-3]. In particular a magnetic inhomogeneity can lead to the creation of triplet superconducting correlations which can penetrate into the ferromagnetic layer over large distances, as predicted for the first time in Ref. [4]. Apart from the ferromagnetic exchange field, superconductivity can also be modified by local stray fields. Those fields might originate from a specific sample geometry or from domains or domain walls (DWs) in close proximity to a superconductor. Magnetic non-uniformities adjacent to a superconductor may therefore lead to the enhancement or reduction of the superconducting condensation energy [5-8]. Among other factors the observation of one or the other behavior depends on the magnetic configuration [6-8], on the size of domains relative to the superconducting coherence length [9], and on the kind of DW present in a ferromagnet [10].

Recent experiments have been devoted to study the influence of DW-generated magnetic stray fields onto superconductivity. Those include thin-film, in-plane [11, 12] and out-of-plane [9] magnetized, hybrid SC-FM structures, as well as patterned SC-FM nanostructures [13, 14]. Several recent investigations of SC-FM hybrid systems studied the interaction of superconductivity with relatively weak ferromagnets such as CuNi-alloys [15, 16] and few works were focused on the interplay of superconductivity and stray magnetic fields originating from strong ferromagnetic materials such as Py ($Ni_{80}Fe_{20}$), Co and Fe [5, 6, 17]. Despite the intense research in this field, the interaction between the DW-induced magnetic field and superconductivity has frequently produced ambiguous results, because of the uncertainty of the external magnetic field range, in which the presumed domain and DWs are being created. Moreover, experiments on FM-SC-FM spin valves reported both standard and inverse switching effects. In particular while in systems including relatively weak ferromagnets such as CuNi alloy, the standard switching effect predicted by the theory [18] was observed [15, 16, 19], some experiments with strong ferromagnetic materials have shown either the standard switching effect [6, 20], the inverse one [7] or both [5, 21]. As pointed out in Ref. [22] the spin-valve effect is measured at temperatures close to the critical temperature where superconducting properties are still weak and the influence of magnetic inhomogeneity, in particular from domains and DWs, should be taken into account for a quantitative description of the effect. However, the spin-valve effect theory considers only the effect of the exchange field [18].

The few experimental works applying the FM-SC-FM geometry for the study of the influence of DW-induced magnetic field effects on superconductivity have often used isotropic ferromagnetic materials without pre-defined easy axis (EA) of magnetization. Those materials do not have a well-defined domain configuration and domain existence regime, so that the experiments could not establish the origin of the observed resistance features unambiguously. In addition, by inserting a superconductor in between the ferromagnetic electrodes the spin accumulation process in the antiparallel orientation of magnetization causes the appearance of additional magnetoresistive effects, which often leads to a masking of the DW-induced response [7, 10, 22].

In the current work we present an experimental study on the influence of stray fields originating from DWs onto the superconducting state of an adjacent Nb layer. In order to avoid effects related to spin-accumulation instead of conventional spin-valve geometry we use a giant magnetoresistance (GMR) sensor capped with a Nb layer. This kind of structure has been previously proposed by Oh et al. [23] for the purpose of studying the superconducting proximity effect in either parallel or antiparallel state of magnetizations of electrodes. Several groups have used this sample geometry to study the interaction of superconductivity and magnetism. However a majority reported difficulties in achieving a clear determination of magnetization states of ferromagnetic electrodes away from the saturation [24]. In our structure the bottom electrode of the GMR sensor is characterized by an uniaxial in-plane magnetic anisotropy and is furthermore coupled ferromagnetically to the top electrode. Ferromagnetic electrodes with in-plane uniaxial magnetic anisotropy exhibit stable single domain states and form multiple domain states only in a narrow and well-defined region of applied external magnetic field. This allows for the clear identification of the field range in which domains and DWs occur and thus magnetic stray fields onto adjacent layers are produced. Due to a properly adjusted ferromagnetic interlayer exchange coupling with the Co-film, we have achieved a uniaxial behavior of the otherwise isotropic Py layer, which enables a similar level of control for the generation of DWs in Py. The combination of two magnetic electrodes with different anisotropy values allows for their independent switching and facilitates the arrangement of their relative magnetizations, which can be either collinear or non-collinear. By probing the GMR response of our sample we are able to track the relative magnetization orientations of the electrodes during transport measurements. So overall, our specific sample design with clearly defined magnetization states and the ability to follow the magnetization reversal accurately by means of measuring the magnetoresistive response only, allows for a clear identification of resistance features that are generated by the DW stray fields in the adjacent superconducting Nb-film.

## II. STRUCTURE

Pseudo-spin-valves were fabricated using DC magnetron sputtering onto HF-etched Si (110) substrates. The specific strategy for the bottom electrode of the spin valve was based on our previous work [25] and the resulting properties of the bottom Co electrode have determined the structural and magnetic properties of the entire sample. The detailed deposition sequence of our sample was as follows: Si-substrate/75nm Ag/50 nm Cr/15 nm Co/4 nm Cu/4 nm Py/100 nm Nb. Buffer layers of Ag and Cr deposited on a single crystal Si(110) substrate create a template for the growth of highly textured hcp (10$\underline{1}$0) Co layers with the c-axis as the magnetically easy axis oriented in the plane of the film [26]. Overall the epitaxial relationship between constituent layers is: Si(110)/Ag(110)/Cr(211)/ Co(10$\underline{1}$0). The epitaxy extends up to the Co layer, leaving Cu and Py layers in their polycrystalline states, which however does not impact the reliability of magnetization state generation as discussed below. All samples discussed in the current manuscript were cut down to a rectangular shape with a size equal to approximately 6 mm by 3 mm.

### III. RESULTS AND DISCUSSION
#### a. Magnetization reversal

Magnetic hysteresis loops were measured using a superconducting quantum interference device-vibrating sample magnetometer (SQUID-VSM) assembly in the temperatures range from 3.0 K up to 10.0 K, which is the temperature range encompassing the superconducting transition of our Nb layers with a critical temperature $T_c$ = 4.07±0.02 K [27].

Overall, the total number of hysteresis loop measurements in this temperature range was 30 for every orientation of the magnetic field. Figure 1 shows the central portion of all 30 hysteresis loops measured along the EA of Co (a) and another 30 loops at a field orientation of 45$^0$ away from the EA (b). Hysteresis loops measured along the EA show abrupt two-step magnetization switching, which results in a square-like shape of the overall curve. The first magnetization switch at ±170 Oe corresponds to the Py layer, while the second switch at ±600 Oe marks the reversal of Co magnetization. The presence of the uniaxial anisotropy assures the creation of collinear parallel and antiparallel states of the two magnetic layers in this case. The nearly perfect reproducibility of the hysteresis loops in the entire temperature range measured here confirms the temperature stability of the magnetization reversal process in the vicinity of the superconducting transition temperature. In contrast to room temperature measurements of these types of samples, for which sample-size magnetization reversal avalanches are observed, the reduced mobility of domains at low temperatures stabilizes multi-domain states in a narrow, but non-vanishing field window around the switching field of each layer, which is found to be temperature independent in the range of interest here.

Figure 1 (b) shows hysteresis loops measured with the applied field oriented $45^0$ away from the EA in the same range of temperatures as in Fig 1 (a). The separate switching of the Py and Co layers is still apparent. However, due to the torque that the applied magnetic field generates in this configuration and the different anisotropy constants of the two layers, their magnetization state orientation is field dependent and not necessarily collinear anymore. This is caused by a tilt of the magnetization in the layers towards its EA and away from the applied field direction below and above the threshold magnetic field where the switching occurs, which also causes the hysteresis loop to have a non-square-like appearance. Importantly, the occurrence of non-collinear magnetization states does not change the fact that the field range of DW existence is very well defined also in this configuration, being limited to the narrow range of applied fields, in which the switching occurs. Similar to the case of EA magnetization reversal, the switching fields of Py and Co are stable and do not change over the measured range of temperatures. Thus overall, our samples allow for the reproducible creation of DWs in a narrow and well-defined magnetic field regime, so that effects of DWs can be clearly separated from direct field effects.

### b. Magnetoresistive response

Transport measurements were performed using a physical property measurement system (PPMS) in the temperature range from 2-10 K and magnetic fields of up to 2 kOe. The specific measurements that we conducted were temperature sweeps for constant applied magnetic field, as well as constant temperature magnetic field cycles. Resistivity measurements were done using a 1mA AC current and utilizing four-point probe geometry. Each magnetotransport measurement, which involved sweeping of the magnetic field was performed after warming up the sample to its normal state and zero-field cooled down to the measurement temperature. Figure 2 (a) shows the temperature-dependent resistance of our structure at zero magnetic field, while Fig. 2 (b) shows several representative resistance versus magnetic field curves for ascending and descending field amplitude with the field oriented along the EA of magnetization, normalized to the zero-field resistance value ($R_0$) at each temperature. The scale bars indicate the percentage change of resistance for each individual panel.

At $T$ = 5K, we observe a very clear GMR effect with a high-resistance plateau for the antiparallel state of magnetization in between the Py and Co layers and a low-resistance plateau for the parallel orientation of their magnetizations. Our structures were explicitly designed for this type of two-plateau magnetoresistance behavior, since it allows for a clear tracking of the magnetization reversal processes in the resistance measurement itself. The observed GMR effect at 5.0 K has only modest amplitude equal to 0.13 %. This relatively low GMR signal can be explained by current shunting through thick buffer layers as well as through the bulk-like Nb layer. Despite its relatively low size, the GMR effect enables the field calibration of the magnetization reversal within the resistance measurement, and thus allows for a precise study of other resistance features in relation to this reference effect. The magnetoresistance curve measured at 4.8 K also shows GMR effect, however slight deformations of GMR plateau are visible. As the temperature is further decreased towards the superconducting transition, the GMR effect becomes less visible and deformations of GMR plateau transform into clear magnetoresistance peaks. These peaks occur exactly at the Py and Co-switching fields and their amplitude continuously increases as we approach the critical temperature of the Nb layer. Also, as the temperature is further decreased along the superconducting transition, the GMR effect becomes less visible being gradually projected onto much higher-in-amplitude magnetoresistance peaks and an overall increasing field-dependent curvature. The parabola-shaped background curvature of magnetoresistance increases as one approaches the critical temperature of Nb. The observed magnetoresistance peaks can be clearly identified up to the point $T$ = 4.0 K, at which Nb becomes superconducting and no field-induced resistance change is visible anymore in the field range used here. The observed resistance peaks coincide with switching of GMR in the proximity to the coercive fields ($H_c$) of the Co and Py layers. Peaks at the lower magnetic field correspond to the switching of Py, while peaks observed at higher fields occur with the switching of the Co layer. The occurrence of resistance peaks close to the coercive fields of the magnetic electrodes and the absence of any other resistance features away from $H_c$ highlights the key role that magnetization reversal processes play in the creation of the observed effect. Since EA reversal of magnetization for both magnetic films involves the formation of stable multi-domain states, the resistance peaks do appear synchronously with the existence of DWs in the magnetic layers, so that it is most appropriate to assume that it is the DW-generated local-stray fields that are responsible for the observed resistance peaks, because these fields will penetrate adjacent films, including the Nb-layer, and thus have the tendency to partially suppress the superconducting phase and cause an increase in the resistance just above the superconducting transition temperature of our Nb-film.

Figure 3 shows field-dependent *dm/dH* curves (line plot in upper panel) and normalized magnetoresistance measurements (data points in lower panels) taken for field orientations along the EA (a) and 45⁰ away from the EA direction (b), respectively. The *dm/dH* curves are calculated from the hysteresis loops presented in Figs. 1 (a) and (b) for their respective field orientations and identify the switching-field regimes of the two magnetic layers and thus the field range, in which DWs occur. In both cases, i.e. for measurements along the EA and 45⁰ away from the EA, we can identify two smaller *dm/dH* peaks, which appear at lower absolute magnetic field amplitude and correspond to switching of the Py layer, as well as two larger peaks which appear at higher magnetic fields and originate from switching of the Co electrode. In the lower panels of Fig. 3 we show the resistance difference $\Delta R(H)=(R_A-R_D)/(R_A+R_D)$, where $R_A$ and $R_D$ denote the ascending and descending branches of the magnetoresistance curve, respectively. For the clarity of view the negative magnetic field branch of $\Delta R(H)$ has been multiplied by (−1). This data representation removes the parabolic background that is seen in Fig. 2 and thus allows for an easier comparison of the magnetometry and magnetoresistance data. It is evident from the experimental data that magnetoresistance peaks are present for both field orientations. Furthermore, the values of the switching field taken as a center position of the *dm/dH* peaks clearly coincide with the positions of the $\Delta R(H)$ peaks, which indicates their origin as a resistive response to the magnetic stray fields generated by the DWs, which penetrate the adjacent Nb layer. DW-induced magnetoresistance peaks for Py have far smaller amplitude in comparison to peaks originating from the Co layer. At first glance, this contradicts the intuitive expectation that the influence of the Py layer as immediate neighbor to the Nb-film is stronger than that of the Co-film. However, this superficial contradiction is actually fully consistent with the explanation of the magnetoresistance peaks as DW-induced, because the total field-generating moment is much smaller for the Py-films as evidenced by the dm/dH-curves, and because the averaged distances of the Py- and Co-layer to the Nb-film, if integrated over the respective layer thicknesses, are not substantially different. In addition, we expect wider DWs to be formed in the low anisotropy Py layer, which in turn decreases the magnetic field generated by its DWs in comparison to the high anisotropy Co layer. If we assume a Neel-type of wall the DW width ($d_W$) can be approximated by [28, 29]: $d_W \approx \pi\sqrt{A/K}$, where $A$ is the micromagnetic exchange stiffness and $K$ is the uniaxial anisotropy energy. For Cobalt the values of $A=1.3\times 10^{-6}$ erg/cm and $K=5\times 10^6$ erg/cm³ lead to 20-nm wide DW. For Py $A$ is of the same order of magnitude, however, it shows a much lower anisotropy, $K = 1 \times 10^3$ erg/cm³, which leads to DW widths of the order of up to 1000 nm and in turn to values of the stray field that are much smaller than those originated by the Co walls.

More detailed analysis of the experimental data shows that the full width at half maximum (FWHM) of *dm/dH* peaks of Co does not fully correspond to FWHM of the corresponding resistance peaks. The FWHM of the *dm/dH* peaks in the EA direction and 45⁰ away from it are equal to approximately 35.0 Oe, while the FWHM of resistance peaks at 4.2 K for the EA and at 45⁰ are equal to 86.0 Oe and 158.0 Oe, respectively. This fact is not surprising, nor does it contradict the DW explanation of the observed resistance effect, since the *dm/dH* peaks represent the field change in the domain area and population of the two collinear magnetization states, while the resistance peaks are caused by the field-dependent nature of the DW network and the magnetic field that they generate in the Nb layer. More importantly the resistance peak widths represent a path-integral property in a laterally non-uniform sample state rather than the volume-averaged resistance value, so that already few DWs can cause substantial current flow suppression. Thus, the micromagnetic structure of the two magnetic layers in the initial or the almost completed magnetization reversal phase is associated with the existence of a sufficient number of DWs to cause an appreciable resistance effect in the Nb-layer that contributes to the resistance peak, while the corresponding magnetization changes are too small to detect. Thus, our DW stray-field explanation would actually predict wider peaks for the resistance measurements than for the magnetometry, further corroborating our explanation. The FWHM of resistance peaks for measurements along the EA is only about half of the ones measured 45⁰ away from the EA. This is expected since as it was discussed before, the DW stability range has an angular dependence and specifically shows a broadening for magnetic fields away from the EA orientation. Therefore, tilting of a sample away from the EA increases the field range in which DWs are formed, so that an overall FWHM increase of the resistance peaks is observed. This enhanced multi-domain stability regime is not necessarily evident in magnetic hysteresis loops (Fig. 1 (b)) since the relevant domains might be a small area fraction of the total magnetic moment of the sample, and thus do not visibly increase the FWHM width of the *dm/dH* peak. Despite this fact, the coincidence of resistance peaks with the *dm/dH* peaks and the angular dependence of the FWHM of the magnetoresistance peaks is fully consistent with DW stray fields as the origin of the here-observed effects.

As already shown on Fig. 2 the amplitude of DW-induced peaks continuously increases as we approach the superconducting critical temperature. This experimental observation suggests that upon decreasing the temperature above $T_c$, the DW-induced effects increase in their relative strength, which is consistent with the overall argumentation for a DW stray-field effect onto the superconducting phase of Nb.

Magnetoresistance data measured at $45^0$ away from the EA also show more complex resistance features, which appear below the saturation field. Those features do not necessarily originate from the interaction of superconductivity with the magnetization state of the ferromagnetic layers, since changing the relative geometry of magnetization and magnetic field away from the EA configuration causes the appearance of non-collinear magnetization states of Py and Co electrodes. These non-collinear states may lead to the appearance of additional secondary resistance features due to anisotropic and/or GMR effects in the ferromagnets alone or due to the appearance of a triplet component of the superconducting condensate as suggested in Refs. [30] and [31]. Due to their complexity the origin of those additional resistance features require more experimental analysis. Importantly our dominating resistance effects, DW-induced resistance peaks, are not obscured by any of those additional features and can be clearly identified due to the special geometry of our sample, which clearly separates the field range in which DWs occur. We have also measured hysteresis loops and magnetoresistance data in the hard-axis magnetization direction (not shown here). While the GMR effect is still present in this configuration we do not observe any DW-induced resistance changes. This is due to the fact that in the case of HA-field orientation, magnetization reversal occurs via coherent rotation only and generally no domain formation should be observed, which is consistent with earlier investigations of these types of in-plane uniaxial magnetic films [32]

Figure 4 shows the amplitude of the DW-induced magnetoresistance as a function of temperature with the magnetic field applied along the EA (triangles) and $45^0$ away from the EA (squares) plotted on the top of resistance versus temperature curve (dashed line). The amplitude of the DW-induced magnetoresistance is higher in the case of the sample oriented $45^0$ away from the EA compared to the case of the EA. This is the consequence of the increased number of DWs upon rotating the magnetic field away from the EA, which is well documented for these types of magnetic films with uniaxial in-plane anisotropy [30]. In both cases, along the EA and $45^0$ away from it, the DW-induced magnetoresistance is clearly enhanced in the immediate vicinity of the superconducting transition of the Nb layer. Its amplitude increases from 0.1 % at 5.0 K up to almost 2.0 % at 4.2 K. No magnetoresistance is observed in the superconducting state of our sample at 4.0 K and below. This mutual dependence of resistance and magnetoresistance confirms the fact that it is the fractional change in the superconducting phase of the Nb layer that is susceptible to the DW-induced stray fields.

## IV SUMMARY

In summary we have investigated the effect of DW stray fields, produced by adjacent FM films, onto the resistive properties of a Nb layer near its superconducting transition. We have used a special sample geometry, which allowed for a clear tracking of the magnetization reversal process and a precise definition of the magnetic field range, in which DWs exist. As we approach the superconducting transition, we observe *R(H)* peaks which surpass the GMR effect in our samples. The occurrence of peaks coincides exactly with the switching of GMR plateaus and switching fields of the magnetic electrodes within the structure. This unambiguously confirms their origin as magnetoresistive response to the DW stray fields. The amplitude of DW-induced magnetoresistance is higher in the case of the $45^0$ field orientation in comparison to the EA alignment with the external field, which is a consequence of the increased number of domains that form in this case. In both cases, EA and $45^0$ away from it, DW-magnetoresistance amplitudes have their maxima in the immediate proximity of the critical temperature of the Nb layer. This indicates that the DW-induced effects scale with the superconducting phase of Nb above the percolation threshold.

Acknowledgements:

We thank F. Lefloch and M. Houzet for valuable discussions. We also acknowledge the Diputacion Foral de Giupuzkoa Ref. 99/11, program Red Guipuzkoana de Ciencia Tecnologia e Innovacion, funding from the Basque Government under Program No. PI2009-17 and UPV/EHU Project IT-366-07, and the Spanish Ministry of Science and Innovation under Projects No. MAT2009-07980 and FIS2011-28851-C02-02.

Fig. 1: (Color online) Set of hysteresis loops measured in the range of temperatures between 3.0 K and 10.0 K along the EA (a) and 45⁰ away from the EA orientation (b). The inset in (b) shows a schematic diagram of our sample.

Fig. 2: (Color online) (a) Resistance versus temperature data showing the superconducting transition at $T_c$ = 4.07±0.02 K, (b) $R/R_0(H)$ curves measured for a set of temperatures between 5.0 K and 4.2 K with the applied field along the EA magnetization direction, $R_0$ indicates the zero magnetic field resistance point, $R_D$ (squares) and $R_A$ (points) denote the descending and ascending field branches of magnetoresistance curve. Scale bars indicate a percentage change of resistance for each individual panel.

Fig. 3: (Color online) $dm/dH$ curves (uppermost panels) calculated from hysteresis loops presented in Fig. 1, and normalized resistance versus magnetic field curves–$\Delta R(H)$ (lower panels) for the case of an EA direction (a) and 45⁰ off the EA (b) orientation of the applied magnetic field. Temperatures as indicated.

Fig. 4: (Color online) Resistance (dashed line) and the amplitude of DW-induced magnetoresistance in the case of magnetic field applied along the EA (triangles) and 45⁰ away from the EA (squares) as a function of temperature.

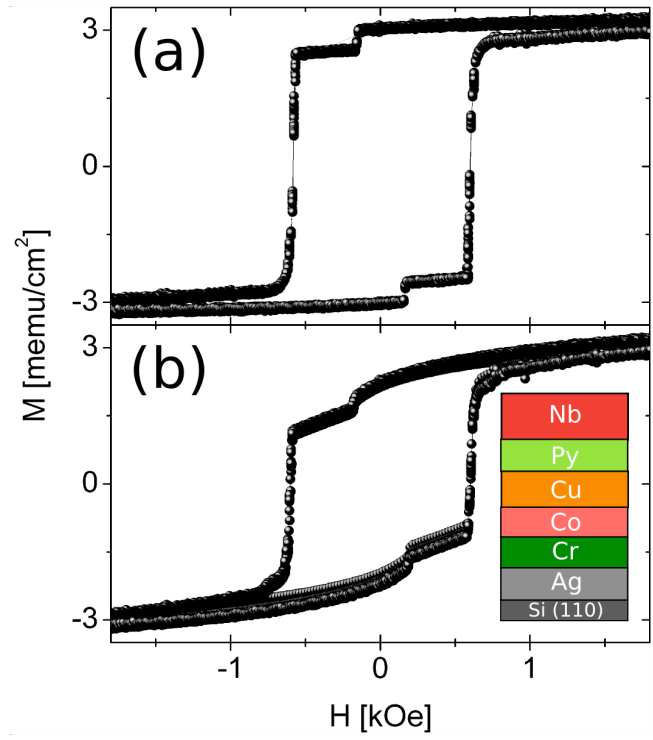

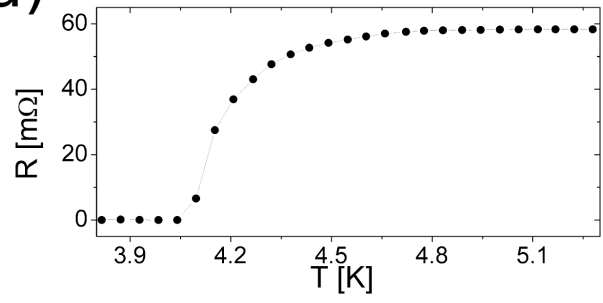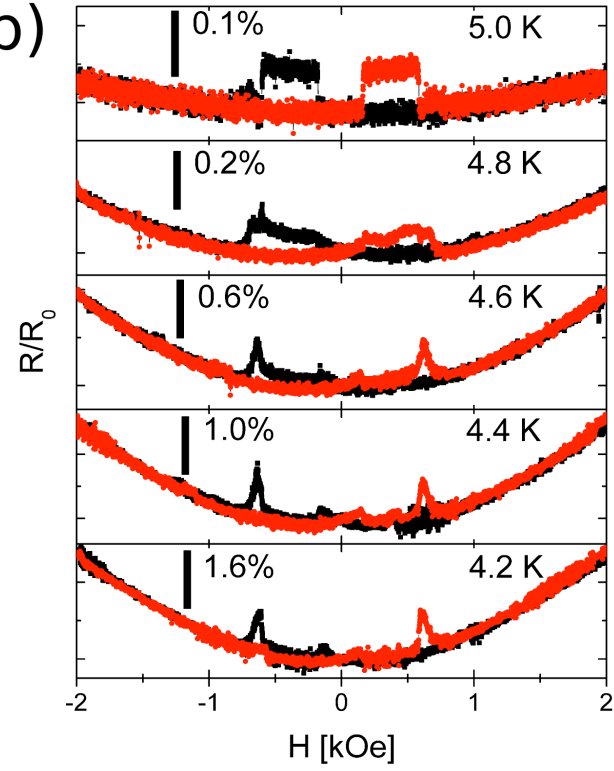

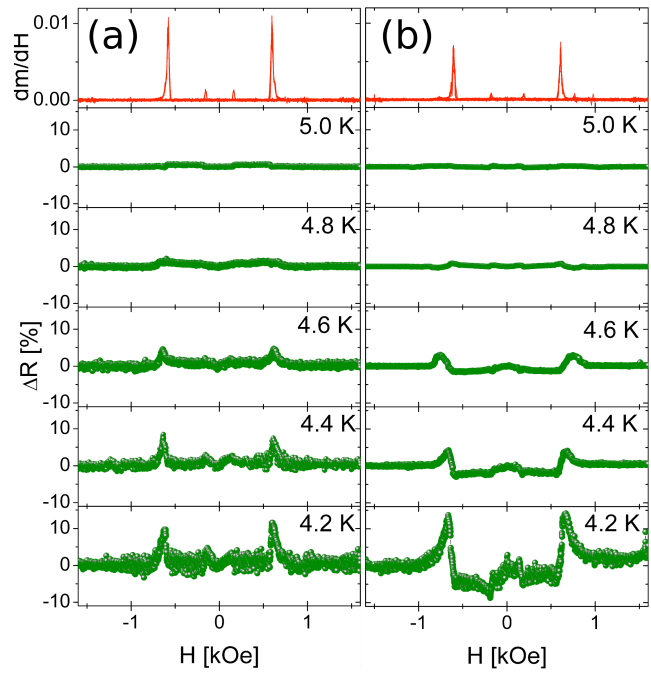

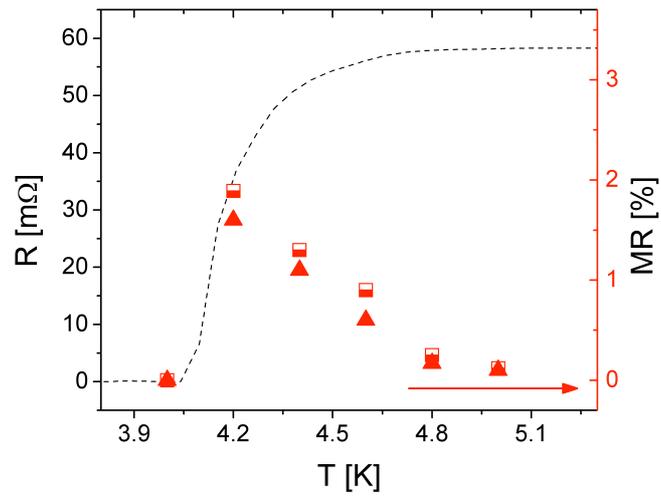